\documentstyle[preprint,aps,epsfig]{revtex}
\tighten
\begin{document}
\draft
\preprint{\parbox[t]{85mm}{Preprint Numbers: \parbox[t]{48mm}
 {ANL-PHY-8240-TH-95 \\ nucl-th/9511012}}}
\title{Electromagnetic Form Factors of Charged and Neutral Kaons}
\author{C. J. Burden\footnotemark[1],
C. D. Roberts\footnotemark[2], and M. J. Thomson\footnotemark[3]}
\address{\footnotemark[1]Department of Theoretical Physics,
Research School of Physical
Sciences and Engineering, Australian National University, Canberra ACT 0200,
Australia\\
\footnotemark[2]Physics Division, Bldg. 203, Argonne National Laboratory,
Argonne IL 60439-4843\\
\footnotemark[3]School of Physics, University of Melbourne, Parkville VIC
3052, Australia}
\date{24/10/95}
\maketitle
\begin{abstract}
The charged and neutral kaon form factors are calculated as a
phenomenological application of the QCD Dyson-Schwinger equations.  The
results are compared with the pion form factor calculated in the same
framework and yield
\mbox{$F_{K^\pm}(Q^2) > F_{\pi^\pm}(Q^2)$} on \mbox{$Q^2\in[0,3]$~GeV$^2$};
and a neutral kaon form factor that is similar in form and magnitude to the
neutron charge form factor.  These results are sensitive to the difference
between the kaon and pion Bethe-Salpeter amplitude and the $u$- and $s$-quark
propagation characteristics.
\end{abstract}
\pacs{PACS NUMBERS: 13.40.Gp, 14.40.Aq, 12.38.Lg,
24.85.+p\vspace*{\baselineskip}  }
%
%
\hspace*{-\parindent}{\bf 1. Introduction}.\hspace*{\parindent}
The kaon is the simplest strangeness-carrying bound state.  Hence studies of
kaon observables and their comparison with analogous pion properties provide
information about $SU_f(3)$-breaking in QCD.  Particularly interesting are
nonperturbative effects, defined as those whose source is the difference
$m_s-m_u$ but which cannot simply be described as a linear response to this
difference.  Examples are the difference between the kaon and pion
Bethe-Salpeter amplitude and that between the $u$- and $s$-quark propagation
characteristics at small and intermediate $k^2$; i.e., $k^2\in[0,1\sim
2]$~GeV$^2$, both of which are generated nonperturbatively.

The electromagnetic form factors of the charged and neutral kaon are
sensitive to these effects and are accessible to experiments at
CEBAF\cite{E93018}.  Herein we report a calculation of a range of pion and
kaon observables via a semi-phenomenological application of the
Dyson-Schwinger equations (DSEs)~\cite{DSErev} in QCD; our primary focus
being the elastic charged and neutral kaon form factors.  In calculating
these form factors we employ a generalised-impulse approximation, in which
the quark propagators (2-point Schwinger functions), meson Bethe-Salpeter
amplitudes (meson-quark vertices) and quark-photon vertices are dressed
quantities whose form follows from nonperturbative, semi-phenomenological DSE
studies in QCD~\cite{DSErev}.  In this way our calculation provides for an
extrapolation of the known large spacelike-$k^2$ behaviour of these Schwinger
functions to the small spacelike-$k^2$ region, where they are unknown and
confinement effects are manifest.  This facilitates an exploration of the
relationship between physical observables and the nonperturbative, infrared
behaviour of these Schwinger functions.

This calculation is a recapitulation, reanalysis and extension of the study
of the pion presented in Ref.~\cite{pion}.  In this approach the quark
propagator has no Lehmann representation and hence may be interpreted as
describing a confined particle since this feature is sufficient to ensure the
absence of quark production thresholds in $S$-matrix elements describing
colour-singlet to singlet transitions.  The quark-photon vertices, which
describe the coupling of a photon to the dressed $u$- and $s$-quarks, follow
from extensive QED studies\cite{DMR94} and satisfy the Ward-Takahashi
identity.  This necessarily entails that the $\gamma$-$K$-$K$ and
$\gamma$-$\pi$-$\pi$ amplitudes are current conserving.  In the chiral limit
the quark-pseudoscalar vertex is completely determined by the scalar part of
the quark self-energy\cite{DS79}, which is a manifestation of Goldstone's
theorem in the DSE approach.  The extension to finite quark masses requires a
minimal modification and preserves Dashen's relation\cite{Dashen69}.

\bigskip

\hspace*{-\parindent}{\bf 2. Generalised Impulse Approximation to
\mbox{\boldmath $F_K(Q^2)$}}.\hspace*{\parindent}
In Euclidean space, with metric \mbox{$\delta_{\mu\nu}={\rm diag}(1,1,1,1)$},
$\gamma_\mu$ = $\gamma_\mu^\dagger$ and
$\{\gamma_\mu,\gamma_\nu\}=2\,\delta_{\mu\nu}$, the generalised impulse
approximation to the $\gamma$-$K^+$-$K^-$ vertex is given by
\begin{eqnarray}
\label{giaus}
\Lambda_\mu^{K^\pm}(p,-q)= \case{2}{3}\Lambda_\mu^u(p,-q)
        + \case{1}{3} \bar\Lambda_\mu^s(p,-q)
\end{eqnarray}
where
\begin{eqnarray}
\label{loop1}
\Lambda_\mu^u(p,-q)  =  2 N_c\,\int\frac{d^4k}{(2\pi)^4}
        {\rm tr}_D\bigl[&&
\bar\Gamma_K(k;p)\,S_u(k+\alpha p)\,i\Gamma_\mu^u(k+\alpha p, k-\beta p +q)
\times\\ \nonumber
&  & S_u(k-\beta p +q)  \,\Gamma_K(k-\beta (p-q);-q)\,S_s(k-\beta p)\bigr],\\
\label{loop2}
\bar\Lambda_\mu^s(p,-q) =
 2 N_c\,\int\frac{d^4k}{(2\pi)^4}\,
         {\rm tr}_D\bigl[ && \Gamma_K(k;p)\,S_s(k+\beta p)
               \,i\Gamma_\mu^s(k+\beta p,k-\alpha p +q)
\times\\ \nonumber
   &&     S_s(k-\alpha p +q)
        \,\bar\Gamma_K(k-\alpha (p-q);-q)\,S_u(k-\alpha p)\bigr].
\end{eqnarray}
Here $q$ is the initial momentum of the kaon, $p-q\equiv Q$ is the photon
momentum and only the Dirac trace remains to be evaluated.  In
Eqs. (\ref{loop1}) and (\ref{loop2}): $S_f$ is the propagator for a quark of
flavour $f=u,s$; $\Gamma_K(p;P)$ is the kaon Bethe-Salpeter amplitude, with
quark-antiquark relative momentum $p$ and centre-of-mass momentum $P$; and
$\Gamma_\mu^{f}$ is the flavour dependent photon-quark vertex.  The parameter
$\alpha$ $[\beta\,=1-\alpha]$ allows for uneven partitioning of the total
momentum between the $u$- and $s$-quark legs connected to the Bethe-Salpeter
amplitude and is fixed by requiring $F_{K^0}(Q^2=0)=0$.  The first of the two
contributions in Eq.~(\ref{giaus}) corresponds to the $u$-quark interacting
with the photon and the $s$-quark acting as a spectator; in the second the
roles are reversed.

If the quark masses are set equal in Eq.~(\ref{giaus}) the generalised
impulse approximation to the $\gamma$-$\pi$-$\pi$ vertex\cite{pion,piloop} is
recovered, provided that $\Gamma_K \rightarrow
\Gamma_\pi$, $f_{K}
\rightarrow f_{\pi}$ and $\alpha = 1/2$, which is required by charge
conjugation
symmetry.

The generalised impulse approximation to the neutral-kaon--photon vertex is
given by
\begin{eqnarray}
\Lambda_\mu^{K^0}(p,-q)= -\case{1}{3}\Lambda_\mu^d(p,-q)
        + \case{1}{3} \bar\Lambda_\mu^s(p,-q)~.
\label{K0FF}
\end{eqnarray}
Herein, except for the electric charge, the $u$ and $d$ quarks are identical
and hence $\Lambda_\mu^d(p,-q)=\Lambda_\mu^u(p,-q)$, defined in
Eq.~(\ref{loop1}).

For elastic scattering  $[p^2=q^2]$ and
\mbox{$
\Lambda_\mu^{K^\pm/K^0}(p,-q) = (p+q)_\mu\,F^{K^\pm/K^0}(Q^2)$},
where $F^{K^\pm/K^0}(Q^2)$ is the electromagnetic form factor.

\medskip

\hspace*{-\parindent}{\bf 2.1}~\underline{Dressed quark propagator}.
\mbox{$S_f = - i \gamma\cdot p\, \sigma_V^f(p^2) + \sigma_S^f(p^2)$}
in Eqs.~(\ref{loop1}) and (\ref{loop2}) can be obtained by solving the quark
Dyson-Schwinger equation\cite{DSErev}.  Realistic, semi-phenomenological
studies provide the basis for the following {\it approximating} algebraic,
model forms for $\sigma_V^f$ and $\sigma_S^f$\cite{pion}:
\begin{eqnarray}
\label{SSM}
\lefteqn{\bar\sigma^f_S(x)  =  C_{\bar m_f}^f\, {\rm e}^{-2x} }\\
& & \nonumber
      +\, \frac{1 - e^{- b^f_1 x}}{b^f_1 x}\,\frac{1 - e^{- b^f_3 x}}{b^f_3
x}\,
        \left( b^f_0 + b^f_2 \frac{1 - e^{- \Lambda x}}{\Lambda\,x}\right)
        + \frac{\bar m_f}{x + \bar m_f^2}
                \left( 1 - e^{- 2\,(x + \bar m_f^2)} \right),\\
\label{SVM}
\bar\sigma^f_V(x) & = & \frac{2 (x+\bar m_f^2) -1
                + e^{-2 (x+\bar m_f^2)}}{2 (x+\bar m_f^2)^2}
                - \bar m_f C_{\bar m_f}^f\, e^{-2 x},
\end{eqnarray}
where $(x=y^2=p^2/\sqrt{2 D})$ and: $\bar\sigma_V(y^2) = 2 D\,\sigma_V(p^2)$;
$\bar\sigma_S(y^2) = \sqrt{2 D}\,\sigma_S(p^2)$; and $\bar m_f$ =
$m_f/\sqrt{2 D}$, with $D$ a mass scale.  The parameters $C^f_{\bar m_f}$,
$\bar m_f$, $b_{1\ldots 3}^f$ are determined either by: 1) fitting a
quark-DSE solution obtained with a realistic gluon propagator; or 2)
performing a $\chi^2$-fit to a range of hadronic observables.  ($\Lambda =
10^{-4}$ is included to decouple the small and large spacelike-$p^2$
behaviour of the $1/p^4$ term, characterised by $b_0^f$ and $b_2^f$.)  We
write the inverse of the quark propagator as
\begin{equation}
\label{ABform}
S_f^{-1}(p) = i\gamma\cdot p\, A^f(p^2) + B^f(p^2)~.
\end{equation}

The quark propagator described by Eqs.~(\ref{SSM})-(\ref{SVM}) is an entire
function in the finite complex-$p^2$ plane and hence does not have a Lehmann
representation.  It therefore admits the interpretation that it describes a
confined particle\cite{DSErev}. The $\sim{\rm e^{-x}}$ form that ensures this
is suggested by the algebraic solution of the model DSE studied in
Ref.~\cite{BRW92}, which employed a confining model gluon propagator and
dressed quark-gluon vertex.

The behaviour of this model form on the spacelike-$p^2$ axis is such that,
neglecting $\ln[p^2]$ corrections associated with the anomalous dimension of
the quark propagator in QCD, it manifests asymptotic freedom.  It has a term
associated with dynamical chiral symmetry breaking ($\sim 1/x^2$) and a term
associated with explicit chiral symmetry breaking $(\sim m/x)$.  Both of
these terms are present in solutions of the quark DSE using a realistic model
gluon propagator~\cite{WKR91}.

\medskip

\hspace*{-\parindent}{\bf 2.2}~\underline{Pseudoscalar Meson Bethe-Salpeter
Amplitude}.
$\Gamma_K$ in Eq.~(\ref{giaus}) is the solution of an homogeneous
Bethe--Salpeter equation (BSE).  Many studies of this BSE suggest strongly
that the amplitude is $\propto \gamma_5$.  Furthermore, in the chiral limit
the pseudoscalar BSE and quark DSE are identical\cite{DS79} and one has a
massless excitation in the pseudoscalar channel with
\mbox{$
\Gamma_{\rm pseudoscalar}(p;P^2=0) =
i\gamma_5\,B_{m=0}(p^2)/f_\pi$}~, where $B_{m=0}(p^2)$ is given in
Eq.~(\ref{ABform}) with $m_f=0$.  This is the realisation of Goldstone's
theorem in the DSE framework; i.e., in the chiral limit Eqs.~(\ref{SSM}) and
(\ref{SVM}) completely determine $\Gamma_{\rm pseudoscalar}$.

Herein, based on these observations, we employ the approximations
\begin{eqnarray}
\Gamma_\pi(p;P^2=-m_\pi^2) & \approx &
        i\gamma_5\,\frac{1}{f_\pi}\, B_{m_u=0}^u(p^2)~,\\
\Gamma_K(p;P^2=-m_K^2) & \approx &
        i\gamma_5\,\frac{1}{f_K}\, B_{m_s=0}^s(p^2)~.
\label{gammaK}
\end{eqnarray}
For the pion this is a good approximation, both pointwise and in terms of the
values obtained for physical observables\cite{FR95}.  For the kaon it is an
exploratory Ansatz, one which need only be accurate as an approximation to
the integrated strength.

\medskip

\hspace*{-\parindent}{\bf 2.3}~\underline{Quark-photon Vertex}.
$\Gamma_\mu^f(p_1,p_2)$ in
Eq.~(\ref{giaus}) satisfies a DSE that describes both strong and
electromagnetic dressing of the quark-photon vertex. Solving this equation is
a difficult problem that has only recently begun to be addressed\cite{MF93}.
However, much progress has been made in developing a realistic Ansatz for
$\Gamma_\mu(p_1,p_2)$~\cite{DMR94}.  The bare vertex, $\Gamma_\mu(p_1,p_2) =
\gamma_\mu$, is inadequate when the fermion 2-point Schwinger function has
momentum dependent dressing because it violates the Ward-Takahashi identity.
In Ref.~\cite{BC80} the following form was proposed
\begin{eqnarray}
\Gamma_{\mu}^{f}(p,k)   = \Sigma_A^f(p^2,k^2)\;\gamma_{\mu}
+ (p+k)_{\mu}\left[\case{1}{2}\gamma\cdot(p+k)\Delta_A^f(p^2,k^2)
- i\Delta_B^f(p^2,k^2) \right]~, \label{VBC}
\end{eqnarray}
with $\Sigma_A^f(p^2,k^2)= [A^f(p^2)+A^f(k^2)]/2$,
$\Delta_A^f(p^2,k^2)= [A^f(p^2)-A^f(k^2)]/[p^2-k^2]$ and
$\Delta_B^f(p^2,k^2)= [B^f(p^2)-B^f(k^2)]/[p^2-k^2]$.
This Ansatz is {\it completely determined} by the dressed quark propagator;
satisfies the Ward-Takahashi identity; has a well defined limit as $p^2\to
k^2$; transforms correctly under $C$, $P$, $T$ and Lorentz transformations;
and reduces to the bare vertex in the manner prescribed by perturbation
theory.  Furthermore, it is relatively simple and hence an ideal form to be
employed in our phenomenological studies.

Using charge conjugation it is straightforward to show that for elastic
scattering one has
\mbox{$(p-q)_\mu\,\Lambda_{\mu}^{K^{\pm,0}}(p,-q) = 0$}
in generalised impulse approximation.  The result
\mbox{$F^{K^{\pm}}(-m_K^2,Q^2=0)= 1 $} follows because the quark-photon
vertex satisfies the Ward identity.

\bigskip

\hspace*{-\parindent}{\bf 3. Calculated Spacelike Kaon Form Factors}.
\hspace*{\parindent}
In the Breit frame:
\mbox{$p=(0,0,\case{1}{2}\,Q,iE_K)$}, \mbox{$q=(0,0,-\case{1}{2}\,Q,iE_K)$},
$E_K= \surd(m_K^2+Q^2/4)$; the calculation of each meson form factor involves
the numerical evaluation of a three-dimensional integral via straightforward
numerical quadrature.

To fix the parameters in the quark propagators we have revised the study of
Ref.~\cite{pion} and refitted the pion observables using the pion mass
formula described in Ref.~\cite{FR95}:
\begin{eqnarray}
\label{pimassform}
m_\pi^2\,f_\pi^2 & = &
\frac{N_c}{2\pi^2}\int_0^\infty\,ds\,s\,
\frac{B^u_{m_u=0}(s)}{B^u_{m\neq 0}(s)}
\left(B^u_{m_u\neq 0}(s)\sigma^u_{S\,m_u=0}(s)
        - B^u_{m_u=0}(s)\sigma^u_{S\,m_u\neq 0}(s)\right)~, \\
\label{fpisq}
\lefteqn{f_\pi^2 = \frac{N_c}{8\pi^2}\int_{0}^{\infty}\,ds\,s\,
\left[B^u_{m_u=0}(s)\right]^2\,
}\\
& &
\left\{  (\sigma^{u}_{V})^2 - 2 \left[\sigma^u_S\sigma^{u\,\prime}_S + s
\sigma^u_{V}\sigma^{u\,\prime}_{V}\right]
  - s \left[\sigma^u_S\sigma^{u\,\prime\prime}_S
        - \left(\sigma^{u\,\prime}_S\right)^2\right]
- s^2 \left[\sigma^u_V\sigma^{u\,\prime\prime}_V
- \left(\sigma^{u\,\prime}_V\right)^2\right]\right\}~.
\nonumber
\end{eqnarray}
with $\sigma^u_{V}\equiv \sigma^u_{V\,m_u\neq 0}$ and $\sigma^u_{S}\equiv
\sigma^u_{S\,m_u\neq 0}$.   This mass formula yields an accurate
estimate of the mass obtained by solving the generalised-ladder approximation
to the pion Bethe-Salpeter equation~\cite{FR95}.

This reanalysis, carried out with the constraint $C^u_{\bar m_u \neq 0} = 0$,
leads to
\begin{equation}
\label{paramVu}
\begin{array}{llll}
C^u_{\bar m_u=0} = 0.121, & \bar m_u = 0.00897, &  &\\
b_0^u = 0.131, & b_1^u = 2.90, & b_2^u = 0.603, & b_3^u= 0.185
\end{array}
\end{equation}
with the mass scale $D=0.160$~GeV$^2$ chosen so as to give $f_\pi =
92.4$~MeV.  Table~\ref{tabres} provides a comparison between calculation and
experiment.

\medskip

\hspace*{-\parindent}{\bf 3.1}~\underline{Differences between $u$- and
$s$-quark propagators}.
Based on the studies of Ref.~\cite{sep} we set
\begin{equation}
\label{paramms}
\bar m_s = 12.5 \,\left( \bar m_u + \bar m_d \right)
\equiv 25\,\bar m_{\rm ave}~,
\end{equation}
where $\bar m_u = \bar m_d$ herein, which is consistent with the theoretical
estimates summarised in Ref.~\cite{PDG94}.  The results we report herein are
qualitatively and quantitatively insensitive to halving or doubling this
ratio.

Realistic DSE studies show that there are differences between the $u$ and $s$
quark propagators that cannot be accounted for simply by changing the mass in
Eqs.~(\ref{SSM}) and (\ref{SVM}).  An example is the vacuum quark condensate,
which, using the model quark propagator defined in
Eqs.~(\ref{SSM})-(\ref{SVM}) and the definition in Ref.~\cite{WKR91}, is given
by\cite{pion}:
\begin{equation}
\langle\bar q^f q^f\rangle_{\mu^2}^{\rm vac} = - (2 D)^{\case{3}{2}}\,
\left( \ln\frac{\mu^2}{\Lambda_{\rm QCD}^2}\right) \,
\frac{3}{4\pi^2}\,\frac{b_0^f}{b_1^f b_3^f}~.
\end{equation}
The current theoretical prejudice\cite{ssbar} is that
\mbox{$\langle \bar s s\rangle^{\rm vac} \sim (0.5 - 0.8)
\langle \bar u u\rangle^{\rm vac}$.}
We have explored the response of our calculated kaon observables to
variations in $\langle\bar s s\rangle$.  We believe that the sensitivity is
too weak to provide a robust, independent estimate of $\langle\bar s
s\rangle$.  However, our calculations favour larger values and hence in the
calculations reported herein we used
\mbox{$
\langle \bar s s\rangle^{\rm vac} =0.8\,\langle \bar u u\rangle^{\rm vac}$},
which was implemented by setting
\begin{equation}
\label{parambs}
\begin{array}{lll}
b_0^s= 0.8\, b_0^u~, & b_1^s = b_1^u ~, & b_3^s = b_3^u~.
\end{array}
\end{equation}

To allow for a minimal residual difference between the $u$ and $s$ quark
propagators we did not allow $C^s_{\bar m_s}$ to vary, simply setting
\begin{equation}
\label{paramC}
C^s_{\bar m_s} = C^u_{\bar m_u} = 0~,
\end{equation}
and allowed variations only in the parameter $b_2^s$.  To provide for a
difference between $\Gamma_\pi$ and $\Gamma_K$ we also allowed
$C^s_{m_s=0}\neq C^u_{m_u=0}$.

Having fixed the $u$-quark parameters in the pion sector we then have a
two-parameter extension of the model to the $s$-quark sector.  These
parameters are fixed by requiring that the model reproduce, as well as
possible, the experimental values for the dimensionless quantities
\mbox{$f_K/f_\pi = 1.22\pm 0.02$},
\mbox{$r_{K^\pm}/r_{\pi^\pm}= 0.88 \pm 0.07$} and
\mbox{$m_K/f_K = 4.37 \pm 0.05$}.

The kaon mass is obtained by solving $\Pi_K(P^2) =0$ where
\begin{eqnarray}
\lefteqn{\Pi_K(P^2)  = 8 N_c\,\int\,\frac{d^4k}{(2\pi)^2}\,
\left(\rule{0mm}{7mm}
B^s_{m_s=0}(k^2)\sigma^s_{S\,m_s=0}(k^2)\, - \right.}\\
& & \nonumber \left.
\left[B^s_{m_s=0}(k^2)\right]^2\,
\left[k_+\cdot k_-\,\sigma^u_{V\,m_u\neq 0}(k_+)\sigma^s_{V\,m_s\neq 0}(k_-)
+ \sigma^u_{S\,m_u\neq 0}(k_+)\sigma^s_{S\,m_s\neq 0}(k_-)\right]
\rule{0mm}{7mm}\right)~,
\end{eqnarray}
with $k_{+} = k + \alpha P$, $k_{-}= k - \beta P$.  For $P^2+m_K^2\simeq
0$, \mbox{$\Pi_K(P^2) \approx f_K^2\,\left(P^2 + m_K^2\right)$} with
\begin{equation}
\label{fKval}
f_K^2 = \left.\frac{d}{dP^2}\,\Pi_K(P^2)\right|_{P^2=-m_K^2}~,
\end{equation}
which also ensures the correct (unit) normalisation of the charged kaon form
factor.  These expressions are straightforward generalisations of
Eqs.~(\ref{pimassform}) and (\ref{fpisq}).  Similar expressions are obtained
in Ref.~\cite{PCTpriv}.

\begin{figure}
  \centering{\
     \epsfig{figure=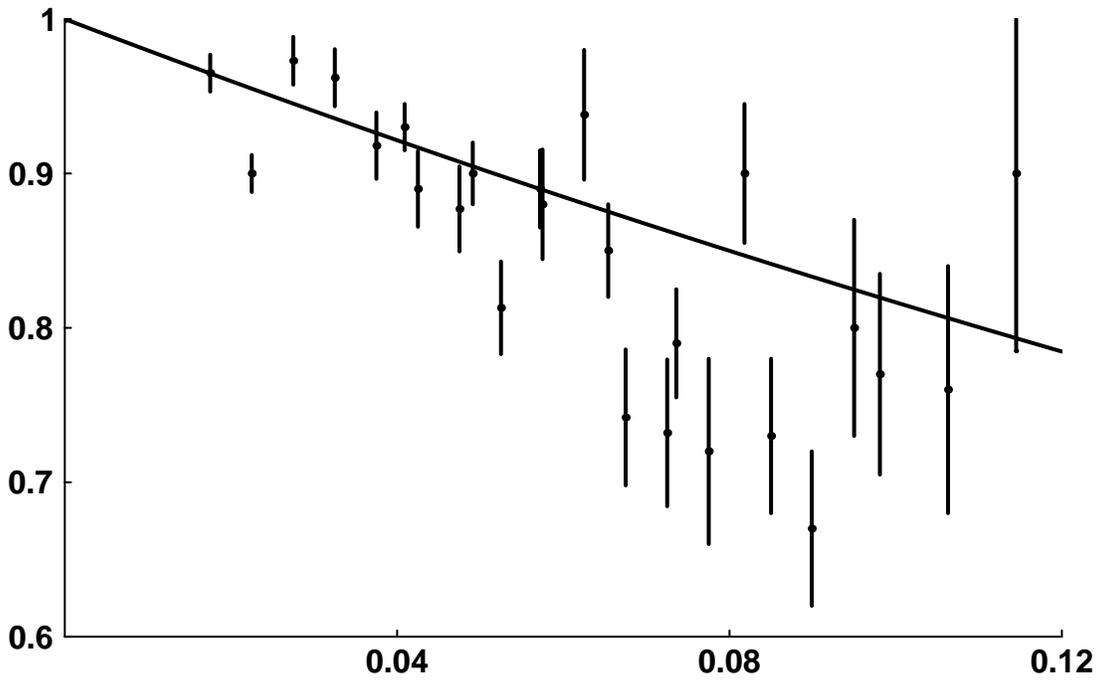,height=20cm,rheight=18cm}  }
\caption{Calculated form of $F^2_{K^\pm}(Q^2)$ compared with the available
data, which is taken from Refs.~\protect\cite{exprkpm} and
\protect\cite{Dally}. ($Q^2$ is measured in GeV$^2$.)
\label{fksqsmq}}
\end{figure}

\begin{figure}
  \centering{\
     \epsfig{figure=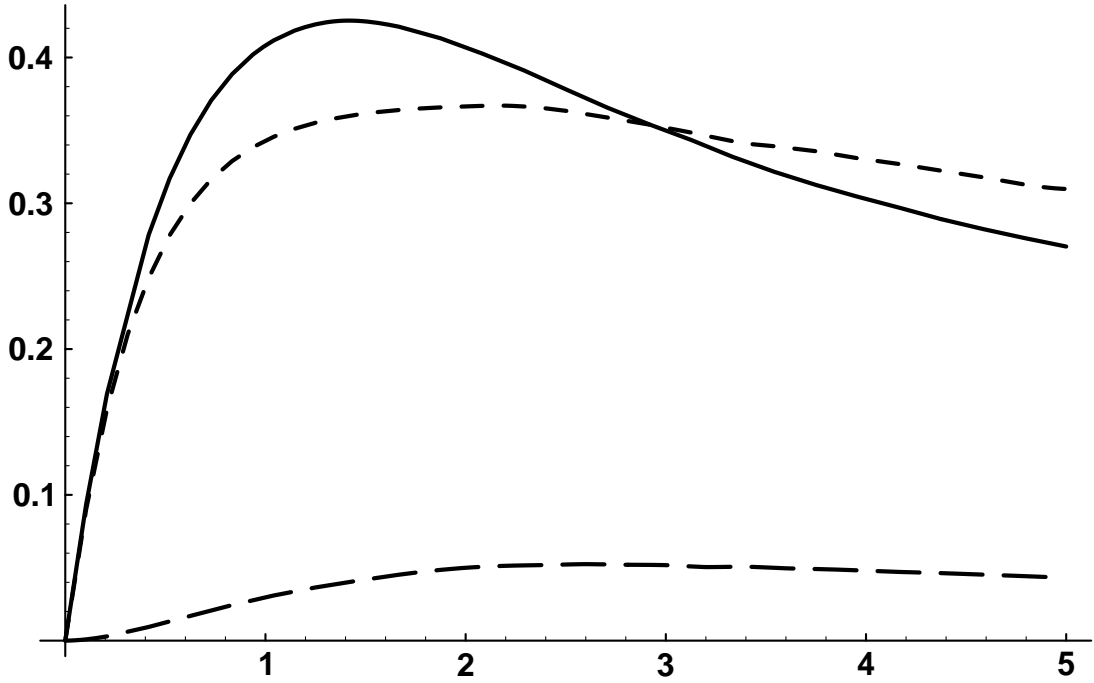,height=20cm,rheight=18cm}  }
\caption{Calculated form factors: $Q^2\,F_{K^\pm}(Q^2)$ - solid line;
$Q^2\,F_{\pi^\pm}(Q^2)$ - short-dashed line; $Q^2\,F_{K^0}(Q^2)$ - long
dashed line.  ($Q^2$ is measured in GeV$^2$.)  The difference between
$F_{K^\pm}(Q^2)$ and $F_{\pi^\pm}(Q^2)$ for $Q^2> 3$~GeV$^2$ is small but is
amplified in this figure because of the multiplication by $Q^2$.
\label{fklgq}}
\end{figure}

Following this procedure we obtain
\begin{equation}
\label{paramVs}
\begin{array}{lll}
C^s_{m_s=0} = 1.69, & b_2^s = 0.74.
\end{array}
\end{equation}
With $\bar m_u \neq \bar m_s$ the requirement \mbox{$F_{K^0}(Q^2=0)=0$}
entails $\alpha=0.49~(\approx 1/2)$.  A comparison between calculation and
experiment is presented in Table~\ref{tabres}.  The calculated form factors
are presented in Figs.~\ref{fksqsmq} and \ref{fklgq}.

The difference between the calculated and measured values of the charge radii
and scattering lengths in Table~\ref{tabres} is a measure of the importance
of final-state, pseudoscalar rescattering interactions and
photon--vector-meson mixing, which are not included in generalised-impulse
approximation\cite{piloop}.  Our calculation suggests that such effects
contribute less than $\sim$ 15\% and become unimportant for $Q^2 >
1$~GeV$^2$.  The fact that the calculated values of $f_K/f_\pi$ and
$r_K/r_\pi$ agree with the experimental values of these ratios suggests that
such effects are no more important for the kaon than for the pion.

\bigskip

\hspace*{-\parindent}{\bf 4. Summary and Discussion}.\hspace*{\parindent}
We find that on the range of $Q^2$ currently accessible to experiment
$F_{K^\pm}(Q^2)>F_\pi(Q^2)$.  This is qualitatively consistent with
Ref.~\cite{BWI95}, however, our calculated results, for both form factors,
are uniformly smaller in magnitude; as is the difference between them.  The
peak in $Q^2 F(Q^2)$, which is most pronounced for the $K^\pm$-meson, is a
signal of quark-antiquark recombination into the final state meson in the
exclusive elastic scattering process.

For $Q^2> 3$~GeV$^2$ we have $F_{K^\pm}(Q^2)<F_\pi(Q^2)$ but the difference
is small and sensitive to the form of the kaon Bethe-Salpeter amplitude.  The
behaviour of $F_{K^\pm}(Q^2)$ at $Q^2>2\sim 3$~GeV$^2$ is influenced by
details of the Ansatz for the kaon Bethe-Salpeter amplitude,
Eq.~(\ref{gammaK}), that are not presently constrained by data.  The results
obtained for $Q^2< 2$~GeV$^2$ are not sensitive to details of our
parametrisation.  We therefore view the results for $Q^2>2$~GeV$^2$ with
caution.

These observations emphasise that measurement of the electromagnetic form
factors is a probe of the bound state structure of the meson; i.e., its
Bethe-Salpeter amplitude.

For the neutral kaon $r^2_{K^0}<0$ and $F_{K^0}(Q^2)$ is similar in form and
magnitude to the charge form factor of the neutron.  The fact that
$F_{K^0}(Q^2)\not\equiv 0$ is a manifestation of the $u$-$s$ mass difference.
We note that charge conjugation symmetry ensures
\mbox{$F_{\pi^0}(Q^2)\equiv 0$}.

Our calculated results are not sensitive to changes in the $m_s$ in the range
$m_s/m_{\rm ave} \in [15,30]$ nor to changes in $\langle\bar s s \rangle$ in
the range $\langle\bar s s \rangle/\langle\bar u u \rangle \in [0.5,1.0]$.
Nevertheless, the requirement that the calculation reproduce known values of
kaon observables does lead to differences between the $u$- and $s$-quark
propagators.  This emphasises that measurement of the form factors is also a
probe of nonperturbatively generated differences between the $u$- and
$s$-quark propagation characteristics.

\bigskip

\hspace*{-\parindent}{\bf Acknowledgments}.\hspace*{\parindent}
We acknowledge useful conversations with A. J. Davies.  This work was
supported by the National Science Foundation under grant no.\ INT92-15223;
the Australian Research Council under grant no.\ S02947481; and the US
Department of Energy, Nuclear Physics Division, under contract number
W-31-109-ENG-38. The calculations described herein were carried out using the
resources of the National Energy Research Supercomputer Center.


\begin{table}
\begin{tabular}{|c|l|l|}
   & Calculated  & Experiment  \\ \hline
  $f_{\pi} \; $    &  ~0.0924 GeV &   ~0.0924 $\pm$ 0.001     \\ \hline
  $f_{K} \; $    &  ~0.113  &   ~0.113 $\pm$ 0.001     \\ \hline
  $m_{\pi} \; $    & ~0.1385  & ~0.1385  \\ \hline
  $m_{K} \; $    & ~0.4936  & ~0.4937  \\ \hline
  $m^{\rm ave}_{1\,{\rm GeV}^2}$ & ~0.0051 & ~0.0075
                \\ \hline
 $m^s_{1\,{\rm GeV}^2}$ & ~0.128 & ~0.1 $\sim$ 0.3 \\ \hline
 $-\langle \bar u u \rangle^{\frac{1}{3}}_{1\,{\rm GeV}^2}$ & ~0.221 &
        ~0.220\\ \hline
 $-\langle \bar s s \rangle^{\frac{1}{3}}_{1\,{\rm GeV}^2}$ & ~0.205 &
        ~0.175 $-$ 0.205\\ \hline
 $r_{\pi^\pm} \;$ & ~0.56 fm & ~0.663 $\pm$ 0.006  \\  \hline
 $r_{K^\pm} \;$ & ~0.49  & ~0.583 $\pm$ 0.041  \\  \hline
 $r_{K^0}^2 \;$ & -0.020 fm$^2$ & -0.054 $\pm$ 0.026  \\  \hline
 $g_{\pi^0\gamma\gamma}\;$ & ~0.505 (dimensionless) & ~0.504 $\pm$
0.019\\\hline
 $F^{3\pi}(4m_\pi^2)\;$ & ~$1.04$ & ~$1$ (Anomaly)\\\hline
 $a_0^0 \;  $ & ~0.17  & ~0.21 $\pm$ 0.01 \\ \hline
 $a_0^2 \;  $ & -0.048 & -0.040 $\pm$ 0.003 \\ \hline
 $a_1^1 \;  $ & ~0.030 & ~0.038 $\pm$ 0.003\\ \hline
 $a_2^0 \; $  & ~0.0015 & ~0.0017 $\pm$ 0.0003\\ \hline
 $a_2^2 \;  $ & -0.00021 & \\ \hline
  $f_{K}/f_\pi \; $    &  ~1.22 &   ~1.22 $\pm$ 0.01     \\ \hline
 $r_{K^\pm} /r_{\pi^\pm}$ & ~0.87  & ~0.88 $\pm$ 0.06 \\
\end{tabular}
\caption{A comparison between the low-energy $\pi$ and $K$
observables calculated using the parameters of Eqs.~(\protect\ref{paramVu})
and (\protect\ref{paramVs}), with the constraints of
Eqs.~(\protect\ref{paramms}), (\protect\ref{parambs}) and
(\protect\ref{paramC}), and their experimental values.  The calculation of
$g_{\pi^0\gamma\gamma}$ is discussed in Ref.~\protect\cite{pion} and
$F^{3\pi}(4m_\pi^2)$ in Ref.~\protect\cite{AR95}.  The quoted
``experimental'' values of $m^{\rm ave}_{1\,{\rm GeV}^2}$, $m^s_{1\,{\rm
GeV}^2}$, $\langle \bar u u \rangle_{1\,{\rm GeV}^2}$ and $\langle \bar s s
\rangle_{1\,{\rm GeV}^2}$ are representative of current theoretical estimates
in other approaches.  Actual experimental values are extracted mainly from
Ref.\protect\cite{PDG94}; $r_\pi$ is taken from Ref.~\protect\cite{Exp86};
$r_{K^\pm}$ from Ref.~\protect\cite{exprkpm};
$r_{K^0}^2$ from Ref.~\protect\cite{exprk0}, and the $\pi$-$\pi$ scattering
lengths, $a^I_J$, are discussed in
Ref.~\protect\cite{RCSI94}.  }
\label{tabres}
\end{table}
\end{document}